\documentclass[journal = nalefd]{achemso}
\usepackage{amsmath}
\usepackage{amssymb}
\usepackage{dcolumn} 
\usepackage{setspace}
\usepackage{array}
\usepackage{graphicx}
\usepackage{booktabs}
\usepackage{gensymb}
\usepackage{multirow}
\usepackage{psfrag}
\usepackage{color}
\usepackage{siunitx}
\usepackage{booktabs}
\usepackage{gensymb}
\usepackage{multirow}
\usepackage{array}
\usepackage{psfrag}
\usepackage{textcomp}
\usepackage{makecell}
\usepackage[colorlinks,linkcolor=blue,anchorcolor=black,citecolor=blue,urlcolor=black]{hyperref}
\nocite{*}

\newcommand{\eqn}[1]{\mbox{Eq.\hspace{1pt}(\ref{#1})}}

\newcommand{\eqtn}[2]{\begin{equation} \label{#1} #2 \end{equation}}

\DeclareMathOperator*{\argmin}{arg\,min}

\def\br{{\mathbf{r}}}

\def\vw{{von Weizs\"{a}cker}}

\graphicspath{ {./figures/} }

\author{Xuecheng Shao} 
\email{xuecheng.shao@rutgers.edu}
\author{Wenhui Mi} 
\email{wenhui.mi@rutgers.edu}
\author{Michele Pavanello}
\email{m.pavanello@rutgers.edu}
\affiliation{Department of Chemistry, Rutgers University, Newark, NJ 07102, USA}
\altaffiliation{Department of Physics, Rutgers University, Newark, NJ 07102, USA}
\title{An Efficient DFT Solver for Nanoscale Simulations and Beyond}

\begin{document}
\maketitle
\begin{abstract} 
	We present the One-orbital Ensemble Self-Consistent Field (OE-SCF) method, an {alternative} orbital-free DFT solver that extends the applicability of DFT to system sizes beyond the nanoscale while retaining the accuracy required to be predictive. OE-SCF is an iterative solver where the (typically computationally expensive) Pauli potential is treated as an external potential and updated after each iteration. Because only up to a dozen iterations are needed to reach convergence, OE-SCF dramatically outperforms current orbital-free DFT solvers. Employing merely a single CPU, we carried out the largest {\it ab initio} simulation for silicon-based materials to date. OE-SCF is able to converge the energy of bulk-cut Si nanoparticles as a function of their diameter up to 16 nm, for the first time reproducing known empirical results. We model polarization and interface charge transfer when a Si slab is sandwiched between two metal slabs where lattice matching mandates a very large slab size. Additionally, OE-SCF opens the door to adopt even more accurate functionals in orbital-free DFT simulations while still tackling systems sizes beyond the nanoscale.
\end{abstract}
\newpage
Since the mid-sixties, scientists have hoped that one day
first-principles device-level and large-scale materials engineering would be
feasible and widely available such that computational models could in part 
or totally replace experiments, reaching a new level of scientific discovery termed Lab-2.0 \cite{Gould_2016}. This futuristic vision 
can only be accomplished if accurate {\it ab initio} quantum mechanical electronic
structure methods are computationally cheap and can model system sizes beyond the nanoscale. 
Density Functional Theory (DFT) \cite{hohe1964,kohn1965}, 
is an excellent candidate method as it can be realized in algorithms that scale linearly with system size either using kinetic energy density functionals \cite{wang2000,witt2018orbital,shao2018large,hung2009accurate} or a combination of appropriate basis sets and approximate eigensolvers \cite{goedecker1999linear,bowl2012,Liou_2020,Sena_2011}. However, when DFT models nanoscale system sizes, the introduced approximations can limit accuracy \cite{Gonzalez_2002,wang1992}. The alternative is to require massive computing infrastructures \cite{weso2015,Nakata_2020}. Thus, the promise of reaching Lab-2.0 has so far been unfulfilled. As it will become clear below, {the proposed One-orbital Ensemble Self-Consistent Field (OE-SCF) method} brings a fresh prospective to these problems. 

To date, there are two kinds of DFT algorithms: Kohn-Sham DFT (KS-DFT) and orbital-free DFT (OF-DFT). KS-DFT is most common, uses a prescription \cite{kohn1965} whereby the lowest $N_e$ eigenvalues (where $N_e$ is the number of electrons) of a one-particle Hamiltonian need to be computed. OF-DFT prescribes to compute just one state \cite{fermi1927} recovering the effect of the other states with pure density functionals \cite{karasiev2012issues}. Overall, on one hand KS-DFT is accurate because it computes the noninteracting kinetic energy functional exactly. It is, however, limited in the system sizes it can approach due to the computational complexity required to compute the many eigenstates \cite{Moussa_2019}. On the other hand, OF-DFT is applicable to large system sizes because the noninteracting kinetic energy is approximated by a pure density functional. However, it is typically limited in the accuracy it achieves \cite{wang2000,witt2018orbital,wang1999,wang1992} unless expensive new-generation functionals are employed \cite{mi2019LMGP,huan2010}.


In this work, we propose a new OF-DFT solver, OE-SCF, that is fast, stable and accurate. Similar to KS-DFT, OE-SCF relies on an iterative solver so that computationally expensive density functionals are evaluated seldom (only once per self-consistent field, SCF, cycle). This allows the employment of new generation kinetic energy functionals, retaining accuracy while removing the need to diagonalize bringing down the computational cost considerably compared to either KS-DFT or conventional OF-DFT. 

In recent years, a new generation of kinetic energy functionals \cite{mi2019LMGP,huan2010,xu2020nonlocal} has brought OF-DFT to tackle semiconductors \cite{mi2018nonlocal,huan2010} and even clusters \cite{mi2019LMGP,xu2020nonlocal} reproducing KS-DFT within a few hundreds of meV/atom. Unfortunately, these functionals coupled with conventional OF-DFT solvers require the evaluation of several tens or hundreds of convolution integrals for each energy evaluation making them computationally expensive \cite{huan2010,PROFESS3,ATLAS,shao2020dftpy}. Because of this, the new generation of kinetic energy functionals \cite{huan2010,mi2019LMGP,xu2020nonlocal} has never been employed in nanoscale simulations.

OE-SCF changes this state of affairs by employing new-generation kinetic energy functionals at no additional computational cost compared to more approximate functionals. It does so by devising an appropriate SCF procedure. This is a paradigm shift for OF-DFT, as it can now probe system sizes that were unthinkable before with no need to sequester massive computational resources or sacrifice the predictivity of the results. This is precisely the type of development that brings the realization of Lab-2.0 a step closer.

\begin{figure}[htp] \centering
	\includegraphics[width=0.9\linewidth]{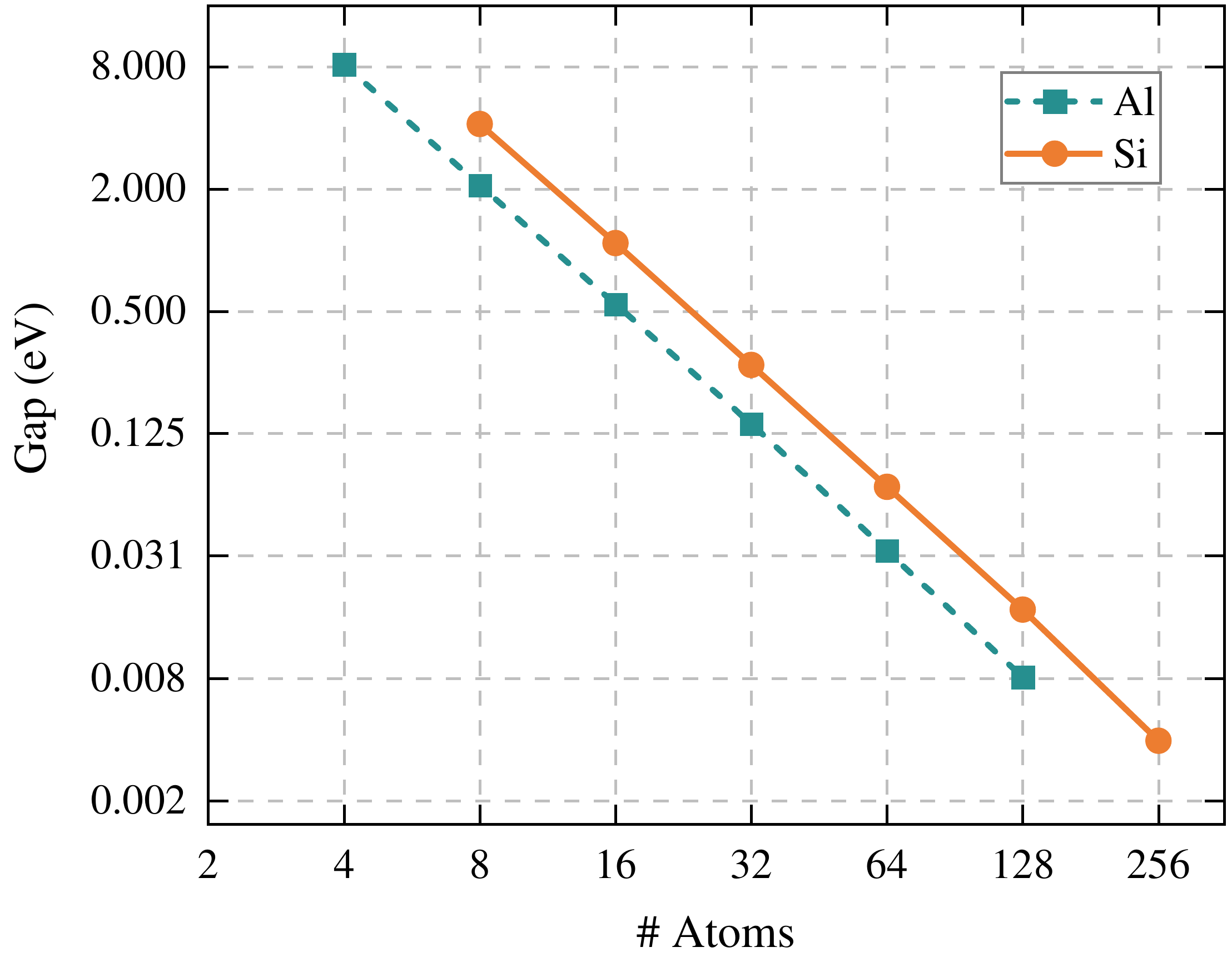}
	\caption{\label{fig:gap} Energy gap
	between the lowest two energy levels associated with the
	eigenvalues of the OF-DFT Hamiltonian in \eqn{eq:ham} for supercells of Al and Si bulk
	systems. The plot is in log scale, displaying a monotonically decreasing gap
	with system size. The LMGP kinetic energy functional\cite{mi2019LMGP} was used.}
\end{figure}

To describe the details of OE-SCF let us consider the OF-DFT Lagrangian, $\mathcal{L}[n]=E[n] - \mu \left(\int n(\br)d\br - N_e\right)$, which is differentiated to find  density functions, $n(\br)$, that minimize the total energy, $E[n]=T_s[n]+E_{Hxc}[n]+\int v_{\rm loc}(\br)n(\br)d\br$ ($T_s$ and $E_{Hxc}$ are the noninteracting kinetic energy and the Hartree-exchange-correlation energies, and $v_{\rm loc}$ is the local external potential), subject to the constraint that the density integrates to a preset number of electrons, $N_e$. Functional differentiation of the Lagrangian leads to the Euler equation of OF-DFT which can be written in a Schr\"odinger-like form,
\eqtn{eq:eig}{\hat h_{OF}\phi(\br) = \mu \phi(\br).} 
Where we introduced the pseudo-wavefunction $\phi(\br) = \sqrt{n(\br)}$, $\mu$ is the chemical potential, and the Hamiltonian is given by
\eqtn{eq:ham}{\hat h_{OF}= -\frac{1}{2} \nabla^2 + v_{\rm Hxc}[n](\br) + v_{\rm Pauli}[n](\br) + v_{\rm loc}(\br),}
where the Pauli potential is the difference between the total noninteracting kinetic energy potential and the potential of the \vw\ functional, $v_{\rm Pauli}[n](\br)=v_{T_s}[n](\br)-v_{\rm vW}[n](\br)$ \cite{karasiev2012issues}, where $T_{\rm vW}[n]=\big\langle \phi \big| -\frac{1}{2} \nabla^2\big| \phi \big\rangle$.

\eqn{eq:eig} derives from the condition of stationarity of the total energy functional. In KS-DFT a similar equation is derived for obtaining the occupied and virtual KS orbitals \cite{kohn1965}. Similarly to KS-DFT, we expect the direct use of \eqn{eq:eig} in an SCF iterative procedure to pose problems of convergence whenever the Hamiltonian (e.g., in \eqn{eq:eig} the Hamiltonian is the one in \eqn{eq:ham}) has dense spectrum. Since 1969, in KS-DFT this is resolved by simply employing ensemble densities {\it via} smearing of the occupation numbers \cite{Slater_1969}. Because smearing requires the computation of potentially very many states (in number that grows linearly with system size), if degeneracy had to appear during the self-consistent solution of \eqn{eq:eig}, smearing the occupations would severely deteriorate OF-DFT's efficiency. 

In the OF-DFT literature \cite{karasiev2012issues,Lehtom_ki_2014,espinosa2015optimizing}, \eqn{eq:eig} could not be numerically solved for any system but those of limited size, containing only a few tens of atoms.\cite{ghosh2016higher,Lehtom_ki_2014,espinosa2015optimizing}
We found the issue to not lie in numerical instabilities as previously thought\cite{Lehtom_ki_2014,espinosa2015optimizing}, but instead it is due to the so-called level-swapping problem arising from degeneracy of frontier states, whereby whenever the energy ordering of occupied and virtual states are swapped from one SCF cycle to the next, the SCF procedure cannot converge \cite{Slater_1969}. We analyze degeneracy in the spectrum of the OF-DFT Hamiltonian of \eqn{eq:ham} in Figure \ref{fig:gap}.

Before introducing our contributions, we should clarify the role of ensemble $N$-representability in OF-DFT simulations. In OF-DFT by far the most common method for solving for the electronic structure is the direct minimization of the energy functional (DEM, hereafter) with respect to the total electron density. Namely,
\eqtn{eq:dem}{n(\br) = \argmin_n\left\{ \mathcal{L}[n] \right\},}
which is insensitive to the features of the spectrum of the OF-DFT Hamiltonian in \eqn{eq:ham}. In those cases when degenerate or close to degenerate levels arise near the frontier orbital (i.e., in OF-DFT the very first orbital) the density resulting from the minimization \eqn{eq:dem} will be an ensemble density. Thus, it is clear that in the presence of degeneracy, solving for the OF-DFT problem with \eqn{eq:eig} can only be done considering ensemble densities -- a procedure that, as stated before, cannot be contemplated because it would require computing many solutions of \eqn{eq:eig}, defeating the purpose of using OF-DFT as an almost linear scaling method.

\begin{table}[htbp]
	\caption{\label{tab:min}Comparing OE-SCF with other OF-DFT solvers. The external potential at iteration $i$ is to be considered constant during the energy minimization. After the minimization is completed, the density-dependent parts of the external potential are updated, as is done in typical SCF procedures. The acronyms are as follows. DEM: direct energy minimization; SCF: self-consistent field; OE-SCF: the newly proposed DFT solver. OE-SCF probes ensemble N-representable densities without the need to compute extra eigenstates through diagonalization.}
	\centering
	\footnotesize 
	\setlength{\tabcolsep}{5pt}{
		\begin{tabular}{llrrr}
		\hline
		\hline
			Method & Functional to minimize & External potential ($v_{\rm ext}$)& Ensemble & SCF \\
			\hline
			DEM & \makecell[l]{$\begin{aligned}
				T_{\rm vW}[n] + E_{Hxc}[n] + T_{\rm Pauli}[n] + \int v_{\rm ext}(\br) n(\br)d\br \end{aligned}$} 	&$v_{\rm loc}(\br)$ & Yes & No \\ 
			
			\hline
			SCF & \makecell[l]{$\begin{aligned}
			T_{\rm vW}[n] + \int v_{\rm ext}(\br) n(\br)d\br \end{aligned}$} &$ v_{Hxc}[n_i](\br) + v_\text{Pauli}[n_i](\br) + v_{\rm loc}(\br)$ & No & Yes \\ 
			
			\hline
			OE-SCF & \makecell[l]{$\begin{aligned}
			T_{\rm vW}[n] + E_{Hxc}[n] + \int v_{\rm ext}(\br) n(\br)d\br \end{aligned}$} &$v_\text{Pauli}[n_i](\br) + v_{\rm loc}(\br)$& Yes & Yes \\ 
					\hline
		\end{tabular}}
\end{table}

Figure \ref{fig:gap} shows the monotonically decreasing gap with increasing system size between the lowest two energy levels of the OF-DFT Hamiltonian of \eqn{eq:ham}. This explains why for small model systems (e.g., number of atoms less than 32) this gap is still large enough for the SCF procedure to converge without problems \cite{Lehtom_ki_2014}. For large system sizes, this gap is too small and the level-swapping problem appears impeding SCF convergence of the OF-DFT problem. As mentioned, a typical way around is to access ensemble $N$-representable densities by slightly smearing the occupations across states within a small window of energy that is still big enough to not fluctuate too much from one SCF cycle to the next \cite{Slater_1969}. Once again, this may appear to not be an option for OF-DFT because smearing would require computing by diagonalization a number of states that grows linearly with system size thereby defeating the purpose of using OF-DFT (i.e., to avoid the $\mathcal{O}(N^3)$ complexity of diagonalizations).

OE-SCF solves the problem by introducing ensemble densities and including the Pauli potential in an iterative fashion. OE-SCF's computational protocol is as follows:
\begin{enumerate}
	\item Given a guess density at iteration $i$, $n_i(\br)$ compute the Pauli potential, $v_{\text{Pauli}}[n_i](\br)$.
	\item Define an auxiliary energy functional and associated Lagrangian. Namely,
		\eqtn{eq:oelag}{\mathcal{L}_\text{OE-SCF}[n,n_i]=T_{vW}[n]+E_{Hxc}[n]+\int \bigg[ v_{\rm loc}(\br)+v_\text{Pauli}[n_i](\br)\bigg]n(\br)d\br}
	\item Minimize $\mathcal{L}_\text{OE-SCF}$ with respect to variations of $n(\br)$. Namely,
		\eqtn{eq:scf}{n_{i+1}=\argmin_{n}\left\{ \mathcal{L}_\text{OE-SCF}[n,n_i] \right\},}
		which, following the reasoning reported after \eqn{eq:dem}, will yield an ensemble density whenever the underlying Hamiltonian features (quasi) degeneracy. For clarity, the underlying Hamiltonian in this case is 
		\eqtn{eq:ham2}{\hat h_{OF} = -\frac{1}{2} \nabla^2 + v_{\rm Hxc}[n](\br) + \underbrace{v_{\rm Pauli}[n_i](\br) + v_{\rm loc}(\br)}_{\text{external potential at iteration $i$}}.}
	\item Repeat (1-3) until convergence (e.g., $\big|E[n_i] - E[n_{i+1}]\big|<10^{-6}$ Hartree for three consecutive cycles).
\end{enumerate}


\begin{table}[htp]
	\caption{\label{tab:pot}Total number of Pauli potential and energy calls during a single-point calculation of bulk Aluminum and Silicon. For OE-SCF the number of calls is equal to the number of SCF cycles needed to reach self-consistency. The LMGP nonlocal kinetic energy functional was employed.}
	\begin{tabular}{rrrrr}
	\hline
	\multirow{2}{*}{\# Atoms}	&\multicolumn{2}{c}{Aluminum} & \multicolumn{2}{c}{Silicon}\\
		\cline{2-5}
		& DEM & OE-SCF & DEM & OE-SCF \\
		\hline
		8	& 41	& 10	& 67	& 20 \\

		32	& 41	& 10	& 67	& 20 \\

		128	& 59	& 13	& 92	& 13 \\

		512	& 200	& 12	& 211	& 13 \\

		2048	& 595	& 8	& 531	& 10\\
		\hline
	\end{tabular}
\end{table}

As is indicated in \eqn{eq:ham2} as well as in Table \ref{tab:min}, in OE-SCF the external potential includes the local (pseudo) potential and the Pauli potential (at iteration $i$) which contains the most expensive energy term in OF-DFT when new-generation kinetic energy functionals are employed (i.e., the nonlocal part). In this way, OE-SCF needs to evaluate the Pauli potential only once for each SCF cycle. A strong motivation for our work is given by a similar procedure \cite{ghosh2016higher} which was shown to be successful for bulk systems when the WGC functional \cite{wang1999orbital} is employed.

Before presenting results from OE-SCF, let us enumerate the computational details. All OF-DFT computations presented in this work are carried out with DFTpy \cite{shao2020dftpy}.  A kinetic energy cutoff of 600 eV is employed in
all OF-DFT calculations, except for interfaces and surfaces where we used a larger cutoff, 1,200 eV. We adopt the following exchange-correlation functionals: local density approximation (LDA) \cite{Perdew1981} for all bulks, clusters and surface energies, and revised Perdew-Burke-Ernzerhof (revPBE) \cite{revPBE} for interfaces. Reference KS-DFT results are calculated with
both the same pseudopotentials as OF-DFT (i.e., local pseudopotentials (LPPs) \cite{Huang_2008,OEPP}, see supporting information document for additional details) as well as nonlocal ultrasoft
pseudopotentials (USPP) \cite{DalCorso_2014}. The calculations with USPPs are
carried out with Quantum-ESPRESSO (QE) \cite{qe} with a 70 Ry kinetic energy cutoff for the wavefunctions. The KS-DFT calculations with LPPs are performed by CASTEP \cite{CASTEP} with a 1000 eV cutoff ensuring that the total energies converge within 1 meV/atom. We adopt the truncated Newton method for the direct energy minimization (DEM) algorithm used to minimize $\mathcal{L}[n]$ and $\mathcal{L}_{OE-SCF}[n,n_i]$. The details of this method can be found in the section 2.3 of Ref.\citenum{ATLAS}.

In Table \ref{tab:pot}, we show that the number of Pauli potential calls for the commonly adopted DEM method grows with system size, while for OE-SCF not only the number of potential calls is much reduced, but is also insensitive to system size. In OE-SCF, the \vw\ and the Hartree-exchange-correlation potentials are invoked as often as in the DEM method.  However, they are computationally affordable if (semi)local xc functionals are adopted (as is most often the case). 

We have studied ways in which density mixing can be used to accellerate the convergence of OE-SCF. We have implemented a density mixing method analogical to Pulay’s DIIS \cite{pula1980}. However, for the systems considered in this work, we notice only a small improvement when density mixing is included. Thus, we decided to not feature results from density mixing-aided OE-SCF. It is possible that the need for mixing is system dependent. Thus, we will keep density mixing in OE-SCF’s implementation in the DFTpy software \cite{shao2020dftpy}.

\begin{figure}[htp]
	\includegraphics[width=0.9\linewidth]{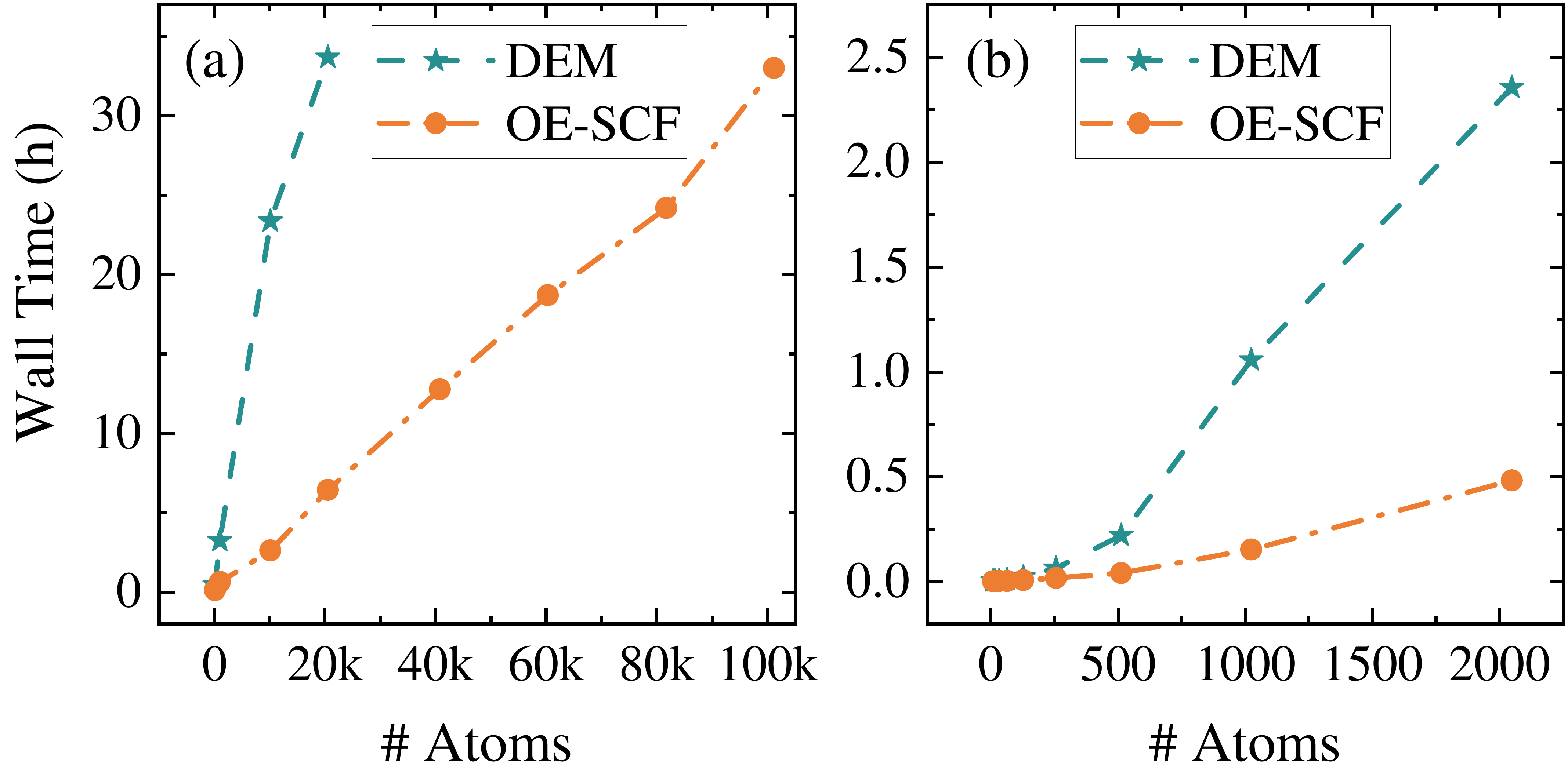}
	\caption{\label{fig:time} Wall times for a single-point calculation of (a) Si
	clusters and (b) Si bulk supercells. DEM is the commonly adopted direct
	energy minimization method in OF-DFT simulations. 
	OE-SCF is the newly proposed DFT solver. Timings
	for Al clusters and bulk supercells are available in the supplementary
	materials.\cite{epaps} The LMGP kinetic energy functional\cite{mi2019LMGP} was used. }
\end{figure}

The number of FFT calls follows the same trend, as is almost constant for varying system sizes as shown in Table S1 and S2. From the timings reported in Figure \ref{fig:time} is clear that OE-SCF is much superior to DEM, cutting the timing down by orders of magnitude in comparison to the current state of the art, maintaining linear scalability with system sizes. 

\begin{figure}[htp]
	\includegraphics[width=0.85\linewidth]{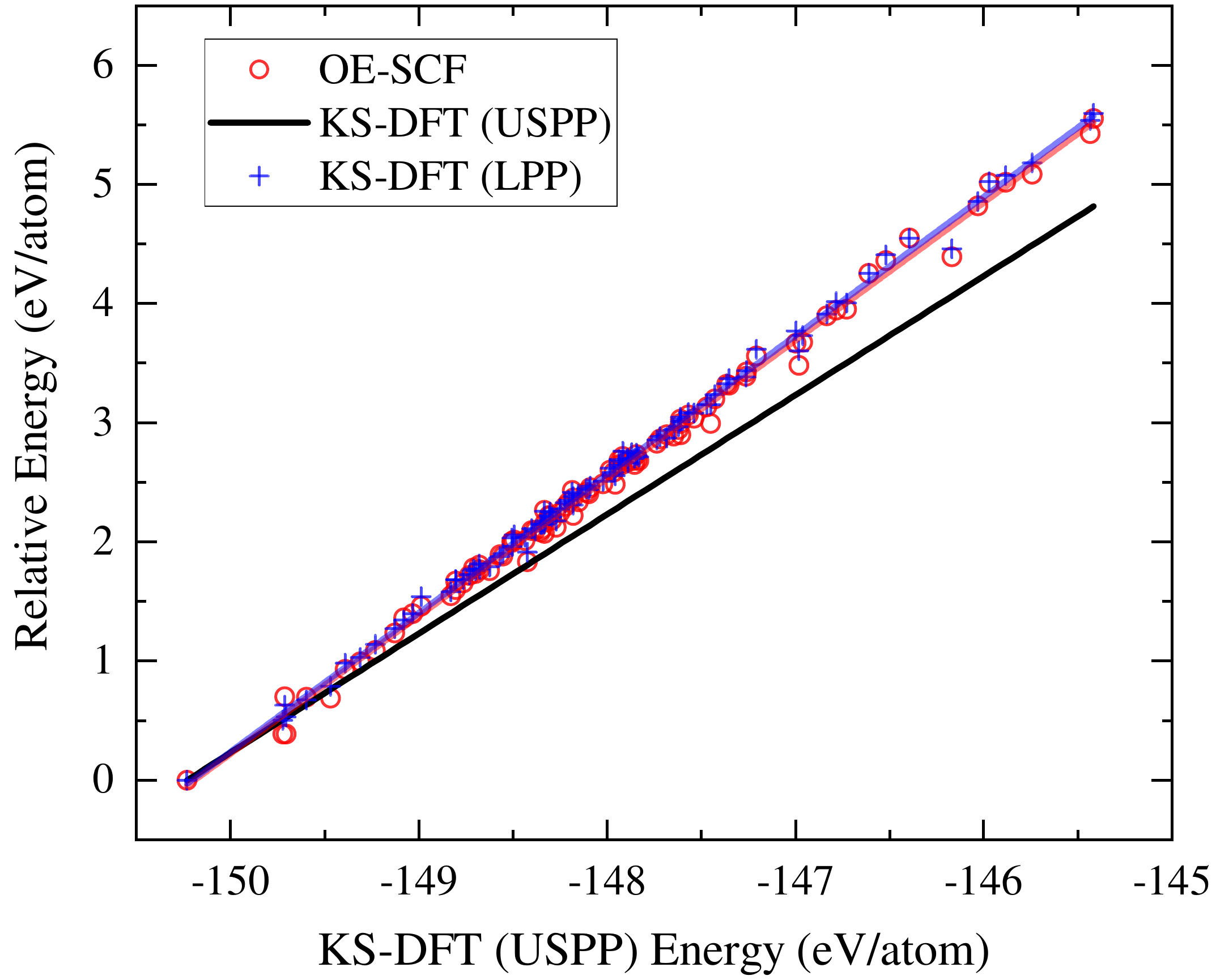}
	\caption{\label{fig:ene} Total energies of 100 random structures of Si$_{200}$ clusters obtained by OE-SCF in comparison with the reference KS-DFT results either with USPPs \cite{DalCorso_2014} or LPPs \cite{Huang_2008}. The latter pseudopotentials are also employed in OE-SCF.}
\end{figure}

OE-SCF reproduces the excellent results of new-generation kinetic energy functionals previously obtained by DEM for clusters and bulk semiconductors \cite{mi2019LMGP,huan2010} indicating that despite the iterative nature, it reaches the variational minimum of the energy functional. In Figure \ref{fig:ene}, we compare OE-SCF total energies (which in this work are always computed with the LMGP nonlocal kinetic energy functional) against KS-DFT for 100 random structures of 200-atom Silicon clusters. The random Si cluster structures are generated by CALYPSO \cite{CALYPSO_CPC,CALYPSO_PRB,CALYPSO_cluster} with the restriction that the minimum interatomic distance is 2.2 \r{A} \cite{mi2019LMGP} and its nearest neighboring periodic images are more than 12 \r{A} apart to ensure the creation of physically feasible structures.

As expected, Figure \ref{fig:ene} as well as Table S2 show that 
OE-SCF's energies lie essentially 
on top of KS-DFT with LPP pseudopotentials and very close to 
KS-DFT with USPP pseudopotentials. To better understand how well OE-SCF ranks the 100 random 
Si cluster structures, we report in Table S2 
the ranking residual standard error (RSE). This is a 
measure of the error from the trendlines plotted in Figure \ref{fig:ene} 
(i.e., deviation from the perfect ranking score). We find that
OE-SCF's RSE is 95/75/66 meV for Si$_{100/200/300}$. We remark that the slopes of the 
ranking trend-line for KS-DFT (LPP) is exactly the same as OE-SCF, and 
its RSE is 68/54/54 meV.
This shows that the majority of OE-SCF's RSE value is not due to the kinetic energy functional employed 
but rather to the inherent differences between LPPs and USPPs. To further quantify the predictivity and accuracy of our approach, we computed surface energies for Si, Mg, and Al and compared against those measured by various experiments. As shown in Table S5, OE-SCF reproduces the experimental results semiquantitatively in line with the expected accuracy of a DFT method.

\begin{figure}[htp]\centering
	\includegraphics[width=0.85\linewidth]{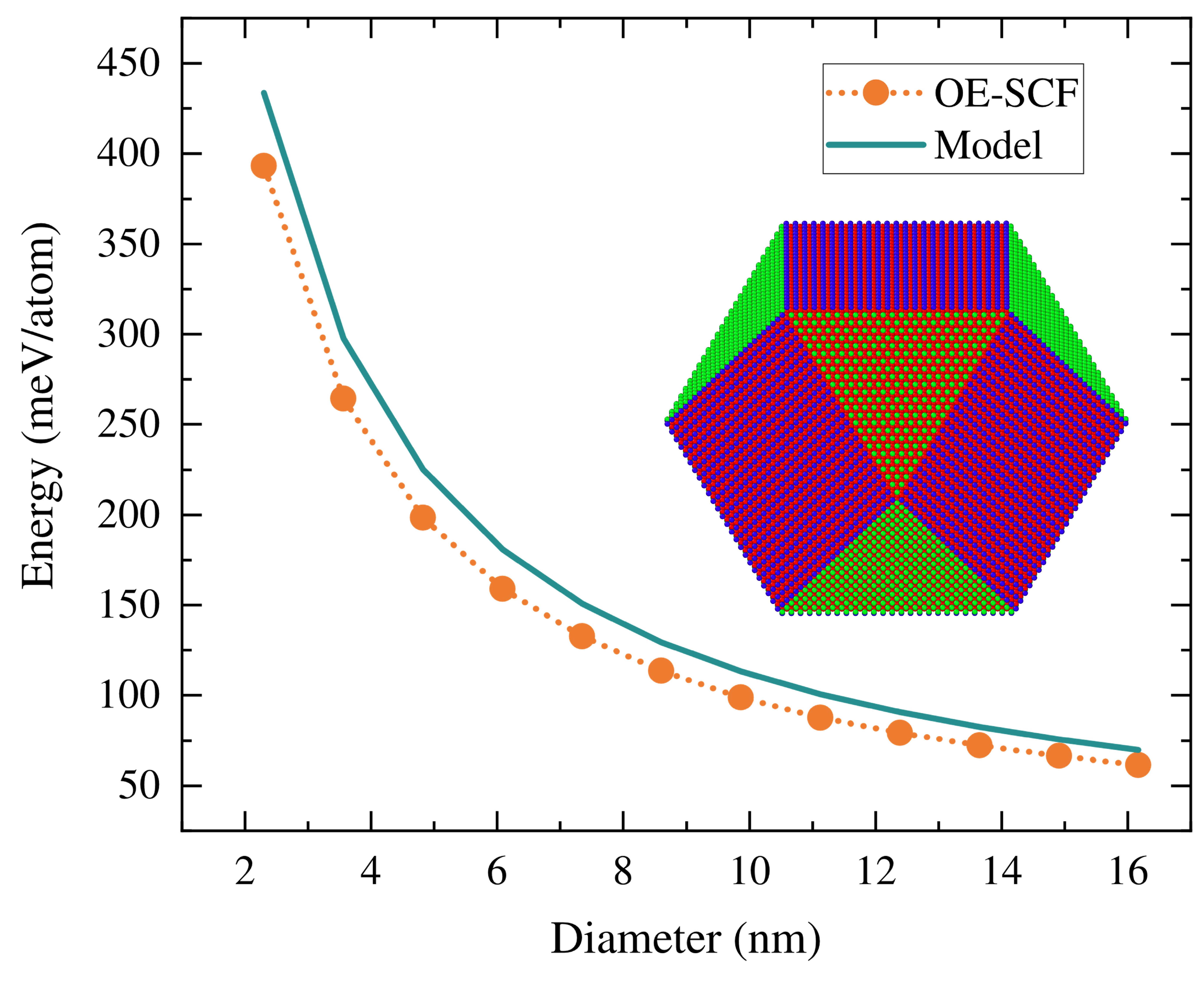}
	\caption{\label{fig:bqd} Total energy per Si atoms versus nanoparticle diameter (dashed orange curve with circles: OE-SCF; solid green curve: empirical formula taken from Ref.\ \citenum{Zhao_2004}). The inset shows the largest Si nanoparticle considered containing 102,501 Si atoms.}
\end{figure}

To showcase the ability of OE-SCF to approach nanoscale systems, we compute the electronic structure of bare and hydrogen-passivated Si nanoparticles (Si-NPs) of several sizes (up to 102,501 Si atoms) and of polyhedral shape. The atomic coordinates of these Si-NPs can be found in the Supporting Information. Figure \ref{fig:bqd} also showcases a plot of the Si-NP's total energy per Si atom converging to the Si bulk value according to a power law. This reproduces previous studies \cite{Zhao_2004,Zhou_2006} up to 7 nm, and extends them all the way up to convergence.
With OE-SCF, not only is it possible to verify the correctness of the empirical power law, but we do so employing a single CPU! In Figure S3 we also show that the energy needed to passivate the Si-NPs used in the study of Figure \ref{fig:bqd} also decays with a similar power law. This indicates that the decay law should also apply to hydrogen passivated Si-NPs and not only to bare Si-NPs.

\begin{figure}[htp]\centering
	\includegraphics[width=0.99\linewidth]{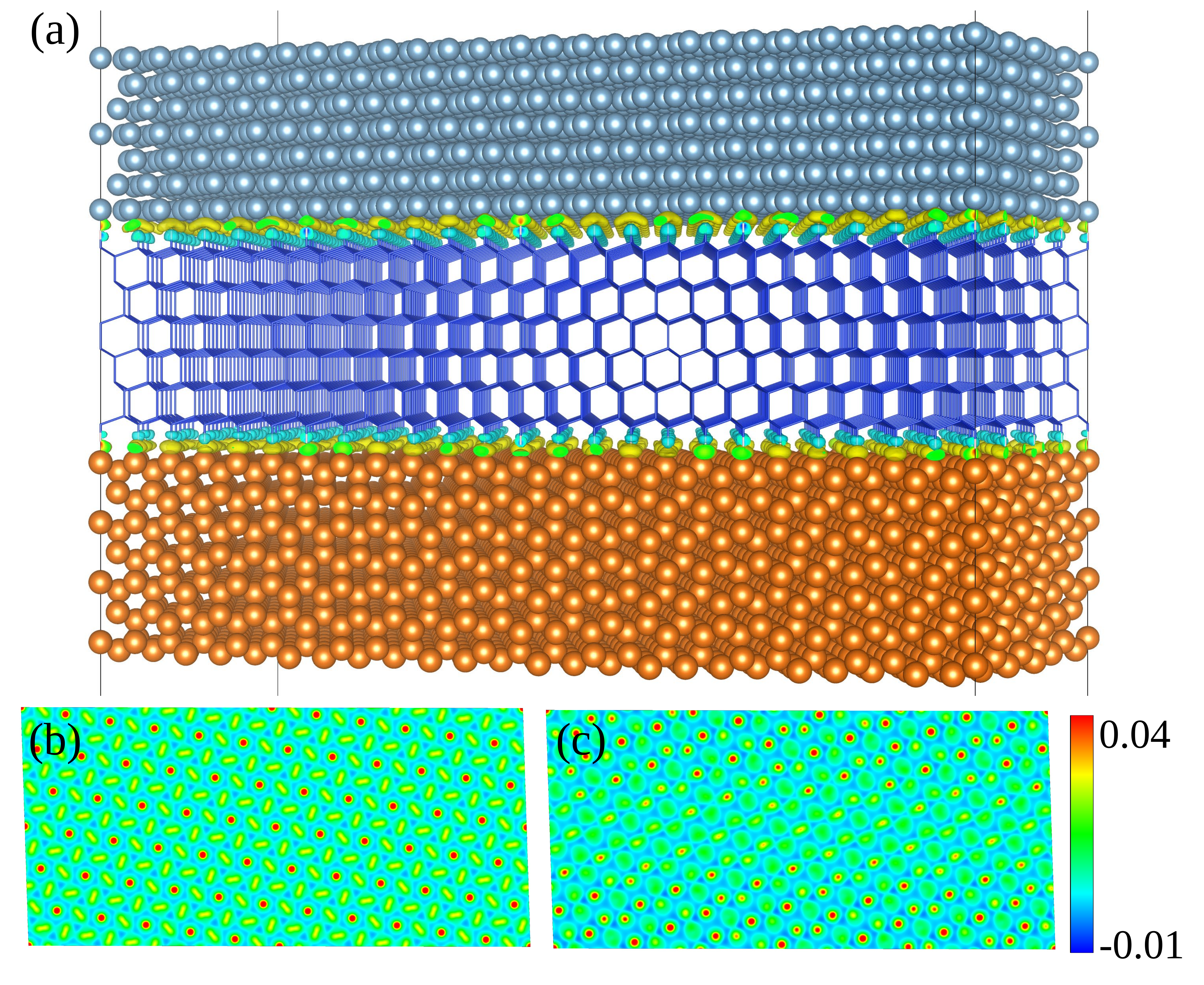}
	\caption{\label{fig:apl} Si-metal interfaces. (a) Al(111) (top), Si(111) (middle) and Mg(100) (bottom, see also Figure S4-S5 for other configurations) which features also the isosurface plot of the polarization density, $n_{\rm pol}(\br)$, defined by subtracting the electron densities of the isolated Mg, Si, and Al slabs from the total electron density of the system for Si-Al and Si-Mg, respectively; (b,c) interface plane cuts depicting iso-values of the polarization density. (b) shows perfectly symmetric interface polarization density while (c) clearly shows that the lattice of Mg(100) and Si(111) differ.}
\end{figure}

As a second example, we inspect the electronic structure of interfaces between unreconstructed bare and hydrogen passivated Si(111) surfaces and Al and Mg metals. We wish to highlight the usefulness of OE-SCF when interfaces are considered. It is often difficult to use DFT methods to describe interfaces because the crystal pacing of the surfaces rarely match \cite{Farmanbar_2016}. With that, large slab sizes, most times outside DFT's realm of applicability, need to be considered to avoid introducing artificial strain to the system.

Figure \ref{fig:apl}a shows the interface considered consisting of 6 layers of Si(111) surface passivated by hydrogen atoms (the bare Si(111) surface is available in Figure S4) with 7 layers of Al(111) on top and 7 layers of Mg(100) at the bottom (the interface with Mg(001) is available in Figure S5). The figure also shows the interface polarization density, $n_{\rm pol}$, (i.e., the charge density difference between the formed interface and the isolated slabs). As we can see from Figures \ref{fig:apl}a, and more clearly in the interface plane cuts in \ref{fig:apl}b and \ref{fig:apl}c, the interface polarization density results in buld up of electrons at the interface. This is consistent with the physics of semiconductor interfaces \cite{FRANCIOSI_1996} and the formation of interface electron gasses \cite{Rotenberg_2003}. We went further and computed the charge of the interface by simply computing the integral $C_i = \frac{1}{2A}\int_{I_i} \left|n_{\rm pol}(\br)\right| d\br $ where $n_{\rm pol} = n - n_{\rm Al}-n_{\rm Mg} - n_{\rm Si}$, $A$ is the surface area, and the integration domains, $I_i$, include either of the interfaces. The values are $C_{\rm SiMg}=1.04 (1.10)$ $e$\AA$^{-2}$ and $C_{\rm SiAl}=1.14 (1.40)$ $e$\AA$^{-2}$ (unpassivated Si interfaces are in parentheses) showing the slight increase in induced polarization when Si has dangling bonds compared to the hydrogen passivated surface. Even though a similar increase is expected as is known for other types of interfaces \cite{Rotenberg_2003}, we will further analyze this system including better sampling of the nuclear configurations in follow-up simulations.

In conclusion, we developed a new DFT solver, called OE-SCF, that leverages
recent advances in OF-DFT development to output a computationally cheap and
accurate {\it ab initio} electronic structure method. The key aspect of OE-SCF
is its ability to still make use of an SCF-like solver capable of sampling ensemble $N$-representable
electron densities while avoiding the diagonalization of the Hamiltonian. We showcase OE-SCF's computational linear scalability with
system sizes well into the hundreds of thousands of atoms for clusters and bulk
systems employing merely a single CPU. Finally, OE-SCF's predictivity is tested computing for the first
time with an {\it ab initio} method the energy decay law with the size of Si nanoparticles
of realistic sizes (up to 16 nm in diameter). We also confirm the decay law which to date had only been tested with ab-initio methods for nanoparticles up to 7 nm in size. Additionally, we showcase examples of Si-metal interfaces where matching crystal pacing is often a show-stopper due to the large simulation cells needed to represent the inherent periodicity of the interface. We compute the interface-induced polarization and interfacial charge transfer predicting slightly larger charge transfer for the interfaces with the unpassivated Si surface.

We have provided important preliminary results indicating that the proposed OE-SCF method will replace the universally-adopted DEM method as the solver of choice in orbital-free DFT software due to its superior stability and computational time to solution.  Simultaneously, because OE-SCF needs to evaluate complex functionals only a dozen times irregardless of the system size, it will enable formulation and deployment of even more complex density functionals than currently available for use in predictive beyond-nanoscale DFT simulations.
\begin{acknowledgement}
This material is based upon work supported by the
National Science Foundation under Grants No.\ CHE-1553993 and OAC-1931473. We
thank the Office of Advanced Research Computing at Rutgers for providing access
to the Amarel cluster. Xuecheng Shao acknowledges the Molecular Sciences Software
Institute for support through a Software Investment Fellowship.
\end{acknowledgement}
\bibliography{paper}
 \end{document}